\documentclass{article}
\usepackage{tikz}
\usepackage{bbm}
\usepackage{booktabs}
\usepackage{threeparttable}
\usepackage{enumitem}

\usepackage{spconf,amsmath,graphicx,hyperref}

\title{Masked Contrastive Pre-Training Improves Music Audio Key Detection}
\name{Ori Yonay, Tracy Hammond, Tianbao Yang}
\address{Department of Computer Science and Engineering, Texas A\&M University}
\begin{document}
%
\maketitle
\begin{abstract}
Self-supervised music foundation models underperform on key detection, which requires pitch-sensitive representations. In this work, we present the first systematic study showing that the design of self-supervised pretraining directly impacts pitch sensitivity, and demonstrate that masked contrastive embeddings uniquely enable state-of-the-art (SOTA) performance in key detection in the supervised setting. First, we discover that linear evaluation after masking-based contrastive pretraining on Mel spectrograms leads to competitive performance on music key detection out of the box. This leads us to train shallow but wide multi-layer perceptrons (MLPs) on features extracted from our base model, leading to SOTA performance without the need for sophisticated data augmentation policies. We further analyze robustness and show empirically that the learned representations naturally encode common augmentations. Our study establishes self-supervised pretraining as an effective approach for pitch-sensitive MIR tasks and provides insights for designing and probing music foundation models. Code and models are available at: \url{https://github.com/echo-cipher/keymyna}.
\end{abstract}
\section{Introduction}
\label{sec:intro}

The key of a musical piece defines its tonal center and harmonic structure, shaping tension, resolution, and overall coherence. Accurate key detection is thus a fundamental task in Music Information Retrieval (MIR), with applications in playlist generation, DJ mixing, and large-scale music similarity search. These use cases demand robust and efficient computational methods.

Traditional approaches extract time-frequency features (e.g., spectrograms or chromagrams) and apply template matching to estimate the most likely key \cite{keyfinder}. While foundational, such methods are genre-specific and sensitive to timbre and instrumentation, limiting generalizability. More recent deep learning approaches infer key directly from spectrograms or chromagrams \cite{endtoendkey, genreagnostic, inceptionkeynet}, achieving strong results but requiring extensive labeled data, genre-specific tuning, and augmentation strategies to generalize broadly.

A central obstacle in advancing key detection is the scarcity of large-scale labeled datasets, as accurate key annotation demands expert knowledge. To address this, we adopt a self-supervised approach that leverages large amounts of unlabeled music to pretrain models that learn meaningful harmonic representations.

In this work, we present KeyMyna, the first systematic study of self-supervised pretraining for musical key detection. We leverage Myna-Vertical as our base model (22M parameters) to effectively capture harmonic relationships without labeled data. Shallow MLP heads (1–16M parameters) are then trained on top of the frozen base model for supervised key detection. Linear evaluation on Myna-Vertical features is already competitive with existing methods, and shallow but wide MLPs yields SOTA results, without complex augmentation pipelines required by prior work.


\section{Related Work}
\label{sec:related-work}

Music key detection has long been a core challenge in MIR. Prior work can be grouped into traditional template matching methods, end-to-end deep learning models, and more recent foundation models. 

\subsection{Traditional Approaches}

Early methods relied on template matching, where time-frequency features such as chromagrams or spectrograms are compared against predefined key templates \cite{keyfinder, pauws2004musical, temperley1999s, noland2007signal, faraldo2016key}. These templates encode pitch class distributions for each key, and similarity measures identify the most likely match. While effective in controlled settings, such approaches are sensitive to timbre, instrumentation, and recording quality \cite{faraldo2016key}, and their dependence on handcrafted features limits generalization to complex harmonic structures. They also exhibit systematic biases, such as favoring certain modes \cite{albrecht2013use}.

\subsection{End-to-End Deep Learning Approaches}

Deep learning introduced models that learn directly from raw or lightly processed audio. CNN-based architectures have been widely used for their ability to capture local spectrogram patterns \cite{endtoendkey}. Inception-style networks \cite{inceptionkeynet} applied to chromagrams leverage multi-scale features for improved accuracy, with InceptionKeyNet \cite{inceptionkeynet} highlighting the benefits of deeper networks and aggressive augmentation. These approaches surpass traditional methods by learning rich representations without heavy feature engineering, but they require large labeled datasets and often rely on genre-specific tuning or augmentation to generalize broadly \cite{genreagnostic}.

\subsection{Foundation Models for Music}

Recent MIR work has focused on large-scale pretrained models such as MERT \cite{MERT}, MULE \cite{MULE}, and MusicFM \cite{MusicFM}, which achieve strong results in tasks like tagging, genre classification, and instrument recognition. However, they underperform in key detection. A major limitation is their emphasis on global attributes (e.g., timbre, rhythm) rather than fine-grained harmonic structure. For instance, CLMR \cite{CLMR} uses pitch shifting in its contrastive pretraining, effectively incentivizing pitch invariance and reducing sensitivity to tonality \cite{JukeMIR}. 

This gap highlights the need for pretrained architectures explicitly designed to capture harmonic information. Our contributions are: (1) demonstrating that masked contrastive embeddings with vertical patches can effectively capture pitch-sensitive structure, (2) systematically evaluating probing and hyperparameter regimes, and (3) empirically showing Myna-Vertical's robustness to data augmentations.

\section{Method}
\label{sec:method}

\begin{figure}[]
\centering
\includegraphics[scale=0.275]{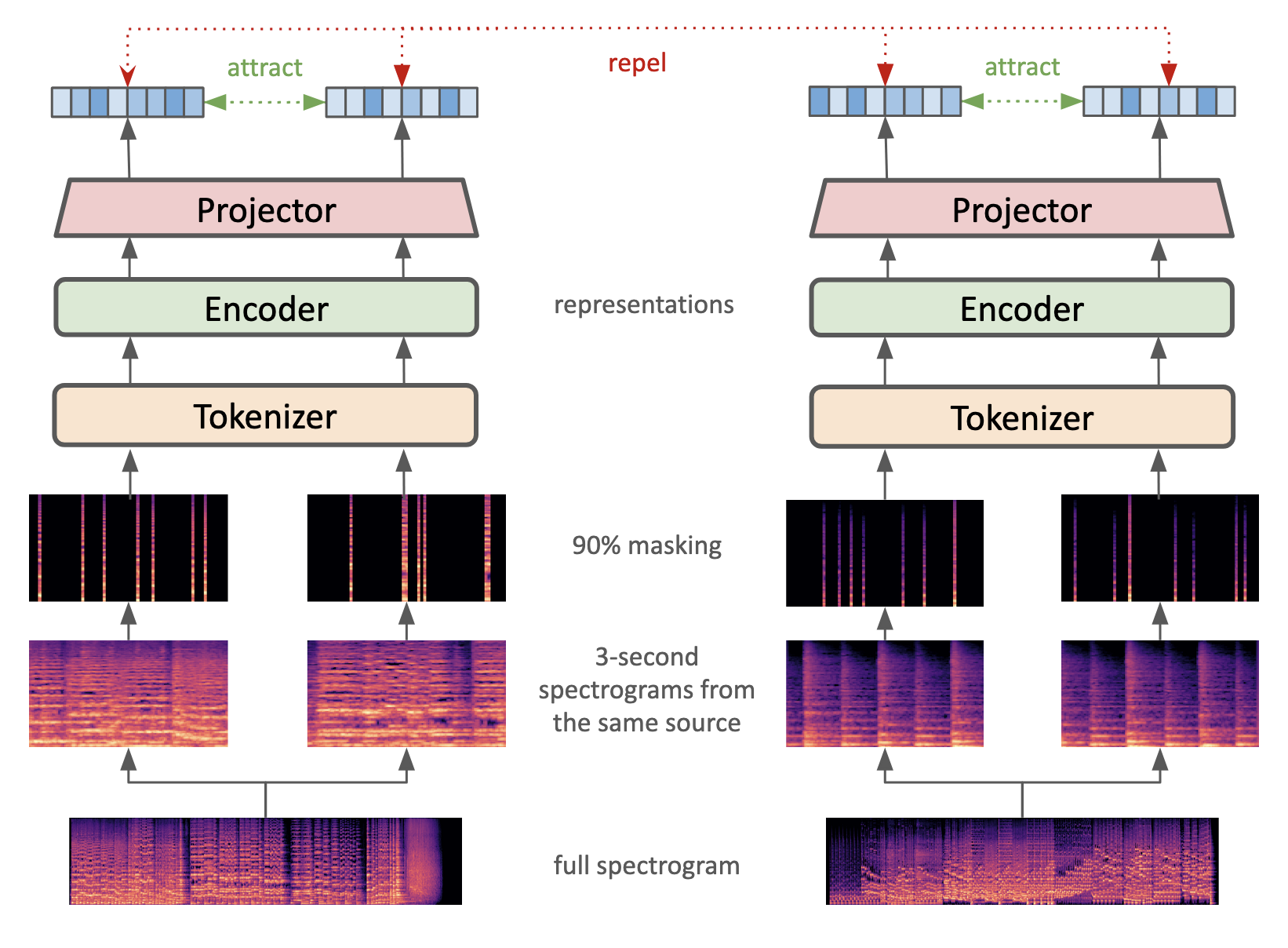}
\caption{Myna's pre-training framework, applied to vertical patches \cite{myna}. 'Encoder' and 'Projector' share weights; the projector is discarded after pretraining, leaving encoder features for downstream experiments. Myna-Vertical refers to the 'Encoder' block.}
\label{fig:visual}
\end{figure}

\subsection{Myna Framework}

Myna is a simple contrastive learning framework that uses token masking as its sole augmentation, originally designed for efficient music representation learning \cite{myna}. It replaces traditional augmentations (e.g., pitch shifting, delay, reverb) with random patch masking (Figure~\ref{fig:visual}). This strategy preserves pitch while improving efficiency, making it well-suited for key detection. 

We use Myna-Vertical as our base model, which was pre-trained to optimize the SimCLR objective \cite{SimCLR}: 

\vspace{-1em}
\begin{equation}
\mathcal{L}_{i} = - \log \frac{\exp \left( \mathrm{sim}(\mathbf{z}_{i}, \mathbf{z}_{j}) / \tau \right)}
{\sum_{k=1}^{2N} \mathbbm{1}_{[k \neq i]} \exp \left( \mathrm{sim}(\mathbf{z}_{i}, \mathbf{z}_{k}) / \tau \right)}
\end{equation}

where $\mathbf{z}_{i}$ and $\mathbf{z}_{j}$ are embeddings of two augmented versions of the same input audio, $\mathrm{sim}(\cdot,\cdot)$ is cosine similarity, $\tau$ is a temperature parameter, and $N$ is the batch size. The loss pulls positive pairs together while pushing all other samples apart. 

Unlike previous work \cite{inceptionkeynet}, which required heavy augmentation, Myna-Vertical achieves robustness from large-scale self-supervised pretraining (roughly three orders of magnitude more data), improving generalization across timbre, instrumentation, and genre. We do not pre-train Myna-Vertical in this work; we use the pretrained checkpoint released by the original Myna authors \cite{myna}.

\subsection{Model Architecture}

Myna-Vertical is a SimpleViT \cite{SimpleViT} model, based on ViT-S/16 \cite{ViT, scalingvit}, which has shown strong performance in both vision and audio domains. Vertical patches ($128 \times 2$) capture all frequency bins at a given time step, encoding harmonic structure while leaving temporal dependencies to be modeled across patches. Square patches, by contrast, blend harmonic and temporal information, making disentanglement harder. 

\subsection{Why Shallow and Wide MLPs?}

Deep MLPs tend to overfit small datasets, while shallow but wide MLPs better balance expressivity and generalization. To regularize, we use high dropout rates, which prevent co-adaptation and encourage robustness. We also evaluate MixUp and find it beneficial for McGill Billboard but detrimental for GiantSteps. We hypothesize that this stems from the greater diversity of the Billboard dataset, where MixUp provides additional robustness.

\subsection{Hyperparameters}

We use 16kHz Mel spectrograms (128 bands, window 2048, hop 512) as Myna-Vertical inputs (chromagram pretraining yielded slightly worse results). Thanks to MLP efficiency, we are able to perform grid search over 160 hyperparameter combinations:

\begin{itemize}[itemsep=2pt,parsep=4pt,topsep=2pt,partopsep=2pt]
\item \textbf{Batch size:} {32, 64, 128, 256, 512}
\item \textbf{Learning rate:} {$10^{-4}$, $3 \times 10^{-4}$, $10^{-3}$, $3 \times 10^{-3}$}
\item \textbf{Weight decay:} {$10^{-5}$, $10^{-4}$, $10^{-3}$, $10^{-2}$}
\item \textbf{MixUp:} [None, ($\alpha=2, \beta=5$)]
\end{itemize}

Batch size, learning rate, and weight decay had little effect on performance. MixUp slightly improved McGill Billboard but hurt GiantSteps. For MLP architecture, we search:

\begin{itemize}[itemsep=2pt,parsep=4pt,topsep=2pt,partopsep=2pt]
\item \textbf{Hidden Layers:} {1, 2}
\item \textbf{Hidden Dimension:} {1024, 2048, 4096, 8192}
\item \textbf{Dropout:} {0.75, 0.9, 0.95, 0.99}
\end{itemize}

We exclude 2-layer, 8192-dim MLPs due to impracticality, leaving 28 settings. 

Myna-Vertical supports variable context lengths due to its attention mechanism. We find 100k samples (roughly 6s) optimal for GiantSteps and 200k (roughly 12s) for McGill Billboard. MixUp ($\alpha=2$, $\beta=5$) is applied only to Billboard. We use a single NVIDIA T4 GPU for MLP training.

\subsection{MLP Configuration}

We extract frozen Myna-Vertical features from 100k- and 200k-sample windows. To expand training data, we apply pitch shifting in [-6,6] semitones. At test time, predictions from each window are averaged.

For GiantSteps, the best model is a two-layer MLP: 384-dim input $\rightarrow$ 4096 units $\rightarrow$ ReLU $\rightarrow$ dropout ($p=0.99$) $\rightarrow$ 4096 units $\rightarrow$ ReLU $\rightarrow$ linear $\rightarrow$ 24 outputs (major/minor keys). For Billboard, the best is shallower: 384 $\rightarrow$ 2048 $\rightarrow$ ReLU $\rightarrow$ dropout ($p=0.75$) $\rightarrow$ linear $\rightarrow$ 24 outputs.

We also train linear models on Myna-Vertical features to demonstrate their representation quality. These models (KeyMyna-\{GS/BB\}-Lin in Table~\ref{tab:key_estimation_results}) contain 9,240 parameters.

\begin{table*}[htbp]
\centering
\begin{threeparttable}
\begin{tabular}{lccccccc}
\toprule
\textbf{Train Dataset} & \textbf{Model} & \textbf{Weighted} & \textbf{Correct} & \textbf{Fifth} & \textbf{Relative} & \textbf{Parallel} & \textbf{Other} \\
\midrule
\multicolumn{7}{c}{\textbf{GiantSteps}} \\
\midrule
Audioset & KeyMyna-GS (ours) & \textbf{75.91} & \textbf{72.02} & 3.48 & 3.64 & 5.30 & 15.56 \\
Audioset & KeyMyna-GS-Lin (ours) & 73.01 & 67.88 & 4.97 & 5.63 & 4.80 & 16.72 \\
Combined\tnote{a} & InceptionKeyNet \cite{inceptionkeynet} & 75.68 & --- & --- & --- & --- & --- \\
Combined\tnote{b} & MERT-95M-Public \cite{MERT} & 72.95 & 67.72 & 4.97 & 5.96 & 4.80 & 16.56 \\
GiantSteps & AllConv \cite{genreagnostic} & 74.60 & 67.90 & 7.00 & 8.10 & 4.10 & 12.90 \\
GiantSteps & ConvKey \cite{endtoendkey} & 74.30 & 67.90 & 6.80 & 7.10 & 4.30 & 13.90 \\
KeyFinder & KeyFinder \cite{keyfinder}\tnote{c} & 59.30 & 45.36 & 20.69 & 6.79 & 7.78 & 19.37 \\
\midrule
\multicolumn{7}{c}{\textbf{McGill Billboard}} \\
\midrule
Audioset & KeyMyna-BB (ours) & 84.35 & \textbf{79.87} & 4.55 & 6.49 & 1.30 & 7.79 \\
Audioset & KeyMyna-BB-Lin (ours) & 81.62 & 76.62 & 4.55 & 7.79 & 1.95 & 9.09 \\
Combined\tnote{b} & MERT-95M-Public \cite{MERT} & 81.30 & 75.97 & 4.55 & 9.74 & 0.65 & 9.09 \\
Billboard & AllConv \cite{genreagnostic} & \textbf{85.10} & \textbf{79.90} & 5.60 & 4.20 & 6.20 & 4.20 \\
Billboard & ConvKey \cite{endtoendkey} & 83.90 & 77.10 & 9.00 & 4.90 & 4.20 & 4.90 \\
\bottomrule
\end{tabular}
\begin{tablenotes}
\item[a] InceptionKeyNet is trained on GiantSteps, McGill Billboard, and a private dataset from data mining.
\item[b] MERT-95M-Public is pre-trained on a mix of publicly-available music audio datasets. 
\item[c] Leaderboard at \url{https://www.cp.jku.at/datasets/giantsteps/}.
\end{tablenotes}
\end{threeparttable}
\caption{Key estimation results on the GiantSteps and McGill Billboard datasets. Results within 0.1\% of SOTA are
bold. InceptionKeyNet is trained on a mix of publicly-available and private datasets and does not have model weights publicly available. KeyMyna refers to shallow MLPs trained on Myna-Vertical features and KeyMyna-\{BB/GS\}Lin refer to linear models trained on these features. KeyMyna is pre-trained on a large unlabeled dataset (that excludes GiantSteps and McGill Billboard). Myna \cite{myna} used a restricted hyperparameter grid without pitch augmentation, yielding lower accuracy than our setup, and therefore not a fair comparison (not shown above).}
\label{tab:key_estimation_results}
\end{table*}

\begin{figure}[t]
    \centering
    \includegraphics[width=0.8\columnwidth]{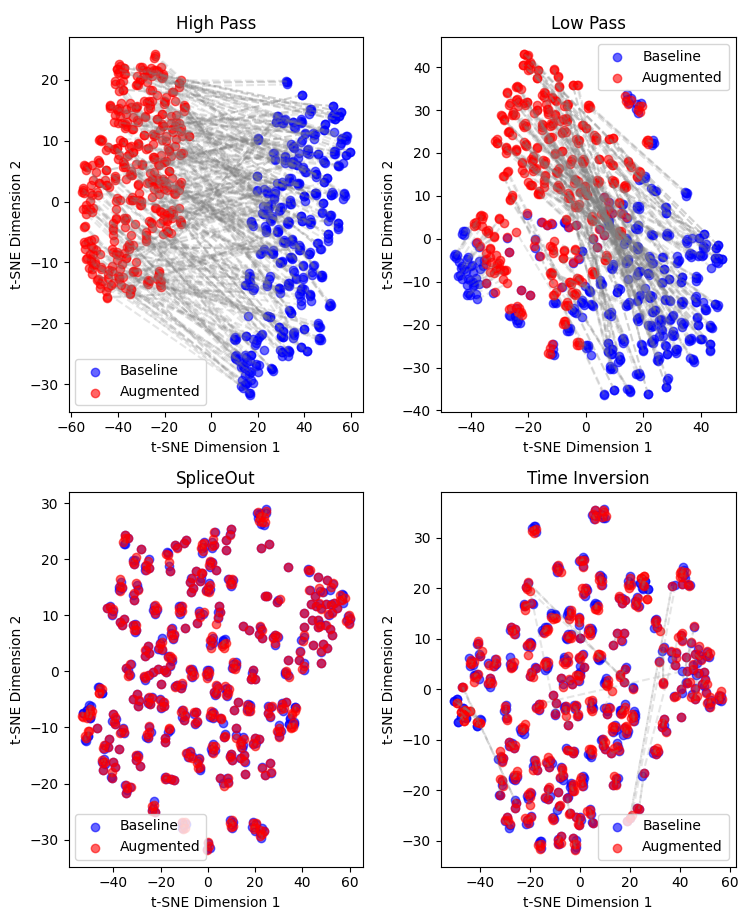}
    \caption{Myna-Vertical is robust to augmentations: we show T-SNE projections of 100 randomly-selected samples from the GTZAN dataset \cite{gtzan} on Myna's embedding space before and after various transformations. Blue points represent the original audio samples and are connected to their augmented counterparts, colored red, with a gray line. The GTZAN dataset was not used to pre-train Myna-Vertical.}
    \label{fig:embedding}
\end{figure}

\subsection{Downstream Datasets and Metrics}

We evaluate on two widely used datasets: GiantSteps and McGill Billboard.

\textbf{GiantSteps}: Provides key annotations for Electronic Dance Music from Beatport user corrections. It includes the GiantSteps MTG Key dataset (1,486 two-minute previews, used for training) and the GiantSteps Key dataset (604 previews with high-confidence labels, used for testing) \cite{giantstepsdataset}.

\textbf{McGill Billboard}: This dataset contains 742 songs from Billboard charts (1958–1991), primarily pop and rock. A subset of 625 songs has annotated tonic and mode labels \cite{endtoendkey}, available at \footnote{http://www.cp.jku.at/people/korzeniowski/bb.zip}

\section{Results}
As shown in Table~\ref{tab:key_estimation_results}, KeyMyna outperforms InceptionKeyNet despite using less data, simpler architecture, and minimal augmentation (only pitch shifting). To compare with other self-supervised methods, we also train MLPs on MERT-95M-Public \cite{MERT} embeddings (previous self-supervised SOTA on GiantSteps). Our results show that masked contrastive pretraining yields more pitch-sensitive embeddings: even a linear probe on Myna-Vertical features surpasses the best MLP on MERT. Linear models on Myna-Vertical embeddings also match prior deep-learning approaches, confirming that large-scale unlabeled pretraining enables strong key detection without heavy augmentation or tuning.

\subsection{Robustness to Augmentations}
Figure~\ref{fig:embedding} shows that Myna-Vertical’s embeddings are robust to common augmentations. We trained linear models to perform augmentations in Myna-Vertical's embedding space and found that most transformations can be accurately approximated linearly. This implies that there are vector directions in Myna's embedding space that correspond to various data augmentations. As a result, it becomes easier for downstream models to learn robustness to these transformations. This result implies that KeyMyna naturally supports performing augmentations in its embedding space. In practice, if a downstream task requires robustness to frequency-specific augmentations (e.g., highpass, lowpass, or bandpass filtering), these transformations can be efficiently applied directly in the embedding space without degrading performance.


\section{Limitations and Future Work}

\subsection{Limitations}
KeyMyna in its current form is only able to track a \textit{global key} - meaning, it is unable to track key modulations within a song, as its predictions are aggregated via averaging. This limitation is manageable for many genres, such as pop, rock, and electronic music, but struggles with pieces that feature key modulations, such as is prevalent in classical music. A potential solution might be to apply a moving average to predictions over time to identify key modulations; we leave this to future work. Additionally, it is unclear how the model will perform in identifying more complex keys (beyond the 24 major and minor keys used in Western music and explored in this work), such as modal keys, microtonal scales, or polytonal structures.

\subsection{Future Work}
Scaling the model size and dataset could improve performance, but this is impractical for CPU-based applications (e.g., DJ software) and offers limited novelty. A more promising direction is fine-tuning the Myna-Vertical model via its MLP adapters. While this work highlights the value of self-supervised pretraining, fine-tuning could better adapt the model to key detection and surpass shallow models trained on embeddings. Another avenue is to use sequence models to aggregate embeddings from multiple song segments; this could better capture temporal dependencies, key modulations, and local harmonic structure. 


\section{Conclusion}
We presented KeyMyna, a systematic study of self-supervised pretraining for music key detection. Using Myna-Vertical, a ViT model trained on Mel spectrograms with vertical patches, we showed that shallow MLPs trained on frozen embeddings achieve state-of-the-art results on key detection benchmarks. Our findings demonstrate that masked contrastive pretraining reduces the need for complex augmentation or dataset-specific tuning while addressing the challenge of limited labeled data.

This study provides the first evidence that self-supervised pretraining can deliver SOTA performance on key detection and offers practical insights for future work on harmonic analysis tasks in MIR.

\vfill
\pagebreak
\bibliographystyle{IEEEbib}
\bibliography{references}

\end{document}